\documentclass
  [letterpaper,11pt,abstract=true,headings=normal,tbtags]{scrartcl} 

\usepackage[letterpaper,margin=1in]{geometry}

\usepackage{mathtools}
\usepackage{dsfont}

\usepackage{calrsfs}

\usepackage[compress]{cite}
\usepackage{enumerate}

\usepackage{graphicx, subfigure}
\usepackage{booktabs}

\let\ALPHABET\mathcal
\let\SET     \boldsymbol
\let\SET     \mathbf
\newcommand\BSC{\mathit{BSC}}

\newcommand\important[1]{{\bfseries#1}}

\def\naturalnumbers{\mathds{N}}
\def\DEFINED{\coloneqq}
\def\EXP{\mathds{E}}
\def\IND{\mathds{1}}
\def\PSP{\Delta}
\def\PR {\mathds{P}}
\def\gobble#1{}
\def\nfrac#1#2{#1/#2}

\def\SEQ#1{\{#1_1,\dots,#1_L\}}
\def\VEC#1{(#1_1,\dots,#1_L)}

\usepackage[standard,amsmath,thmmarks] {ntheorem}
\newtheorem* {problem}       {Open Problem}

\begin{document}

\title {Opportunistic capacity and error exponent regions for compound channel with feedback}

\author{Aditya Mahajan and Sekhar Tatikonda}%

\maketitle

\begin{abstract}
  Variable length communication over a compound
  channel with feedback is considered. 
  Traditionally, capacity of a compound channel without feedback is defined as the maximum rate
  that is determined before the start of communication such that communication
  is reliable. This traditional definition is pessimistic. In the presence of
  feedback, an opportunistic definition is given. Capacity is defined as the maximum rate
  that is determined at the end of communication such that communication is
  reliable. Thus, the transmission rate can adapt to the realized channel. 
  Under this definition, feedback communication over a compound
  channel is conceptually similar to multi-terminal communication. Transmission
  rate is a vector rather than a scalar; channel capacity is a region rather
  than a scalar; error exponent is a region rather than a scalar. In this paper,
  variable length communication over a compound channel with feedback is
  formulated, its opportunistic capacity region is characterized, and lower
  bounds for its error exponent region are provided.
\end{abstract}

\section {Introduction}

The compound channel, first considered by Wolfowitz~\cite{Wolfowitz:1959} and
Blackwell \emph{et.\ al.}~\cite{BlackwellBreimanThomasian:1959}, is one of the
simplest extensions of the DMC (discrete memoryless channel). In a compound
channel, the channel transition matrix $Q_\circ$ belongs to a family $\ALPHABET
Q$ that is defined over a common discrete input and discrete output alphabets $\ALPHABET X$ and
$\ALPHABET Y$. The transmitter and the receiver know the \emph{compound family}
$\ALPHABET Q$ but do not know the \emph{realized channel} $Q_\circ$; the
realized channel $Q_\circ$ does not change with time. We are
interested in characterizing the error exponents of a compound channel 
used with feedback. For that purpose, we define a new notion of the capacity of
the compound channel with feedback. 

There have been comprehensive investigations on the capacity of compound
channels, used both with and without feedback. In addition, there is some work
on characterizing the error exponent of compound channels used with feedback. We
briefly summarize the existing work below, focusing on \emph{finite compound
families} $\ALPHABET Q = \{Q_1, \dots, Q_L\}$. 

Given a coding scheme $S^{(n)}$ defined over a compound family $\ALPHABET Q$,
let $P^{(n)}_\ell$ and $R^{(n)}_\ell$ denote the probability of error and
transmission rate when the realized channel $Q_\circ$ is $Q_\ell$, $\ell = 1,
\dots, L$. The general notion of capacity of a compound channel is as follows:
a rate $R$ is said to be \emph{achievable} if $\forall
\varepsilon > 0$, $\exists$ a sequence $S^{(n)}$ of coding schemes such that
$P^{(n)}_\ell < \varepsilon$ and $R^{(n)}_\ell > R - \varepsilon$, $\ell = 1,
\dots, L$. Then, the capacity is the supremum of all achievable rates. This same
notion applies when the channel is used without or with feedback (the difference
being in the choice of coding schemes $S^{(n)}$). 

When the compound channel is used without feedback, the capacity is given by
(see~\cite{Wolfowitz:1964})
\begin{equation}
  \ALPHABET C_{NF}(\ALPHABET Q) = \max_{P \in \PSP(\ALPHABET X)} \inf_{Q \in \ALPHABET Q} 
  I(P,Q)
\end{equation}
where $\PSP(\ALPHABET X)$ is the space of probability distributions on input
alphabet $\ALPHABET X$ and 
\begin{equation*}
  I(P,Q) = \sum_{x \in \ALPHABET X} \sum_{y \in \ALPHABET Y} P(x) 
  \log \frac{ Q(y|x) } {\sum_{x' \in \ALPHABET X} Q(y|x')P(x')}
\end{equation*}
is the mutual information between the input and output of a channel with input
distribution $P$ and channel transition matrix~$Q$. 
%
%
When the compound channel is used with feedback, the capacity is given by
(see~\cite{ShraderPermuter:2009})
\begin{equation} \label{eq:capacity-feedback-compound}
  \ALPHABET C_F(\ALPHABET Q) = 
  \inf_{Q \in \ALPHABET Q} \max_{P \in \PSP(\ALPHABET X)} I(P,Q) 
\end{equation}
These and other variations of the compound channel are surveyed
in~\cite{LapidothNarayan:1998}.  


The above notion of capacity is pessimistic. It quantifies the maximum rate
\emph{determined before the start of transmission} such that communication is
reliable \emph{over every realized channel $Q_\circ$}. An opportunistic
definition of feedback is possible in the presence of feedback.

For many applications, network traffic is backlogged and a rate guarantee before
the start of transmission is not critical. Rather, we want to communicate at the
maximum rate while ensuring that communication is reliable for \emph{the
realized channel $Q_{\circ}$} (even though $Q_\circ$ is not known to the
transmitter or the receiver before the start of transmission). 
In particular, instead of modeling achievable rate as a scalar value $R$ that is
guaranteed before the start of communication, we model achievable rate as a
vector $\VEC {R}$ such that the rate of communication is $R_\ell$ when the
realized channel is $Q_\ell$. In addition, communication is reliable for every realized
channel. More precisely, we say that a rate vector $\VEC {R}$ is
\emph{opportunistically achievable} if $\forall \varepsilon > 0$, $\exists$ a
sequence $S^{(n)}$ of coding schemes such that $P^{(n)}_\ell < \varepsilon$ and
$R^{(n)}_\ell > R_\ell - \varepsilon$, $\ell = 1, \dots, L$. We define
the union of all opportunistically achievable rates as the \emph{opportunistic
capacity region} $\ALPHABET C_{OF}(\ALPHABET Q)$, \emph{i.e.},
\begin{equation} \label{eq:capacity-feedback}
  \ALPHABET C_{OF}(\ALPHABET Q) = 
  \big\{ \VEC R : \VEC R \text{ is opportunistically achievable} \big\}.
\end{equation}
We formally define opportunistically achievable rates and opportunistic capacity in
Section~\ref{sec:model}.

Let $C_{\ell}$ denote the capacity of DMC $Q_\ell$, $\ell = 1,\dots,L$. Then,
it is straight forward to show (see Corollary~\ref{cor:capacity}) that
the opportunistic capacity region is given by a hyper-rectangle
\begin{equation*}
  \ALPHABET C_{OF}(\ALPHABET Q) = \big\{ \VEC R : 0 \le R_\ell < C_{\ell}, \ 
  \ell=1,\dots,L \big\},
\end{equation*}
which is determined by just its upper corner $(C_{1}, \dots, C_{L})$. Thus, the
capacity region $\ALPHABET C_{OF}(\ALPHABET Q)$ is equivalent to the
\emph{capacity vector} $\ALPHABET C_{\ALPHABET Q} \DEFINED (C_1, \dots, C_L)$.

In this paper, we consider variable length coding schemes. For a
sequence $\{S^{(n)}\}$ of coding schemes that (opportunistically) achieves a rate
vector $\VEC {R}$, we define the error exponent vector $\VEC E$ 
as
\begin{equation*}
  E_\ell = \lim_{n \to \infty} \frac {- \log P^{(n)}_\ell } 
                                     { \EXP_\ell[ \tau^{(n)} ] }
\end{equation*}
where $\EXP_\ell[\tau^{(n)}]$ is the expected length of the coding scheme
$S^{(n)}$ when the realized channel is $Q_\ell$. The union of all
achievable error exponent vectors is defined as the \emph{error exponent region} (EER) at
rate $\VEC {R}$ and denoted by $\ALPHABET E \VEC R$. The formal definition is presented in Section~\ref{sec:model}.

%
%
Consider a DMC~$Q$ used with feedback. Let $C_Q$ denote its capacity. 
The error exponent of variable length
coding scheme at rate $R < C_Q$ is given by (see~\cite{Burnashev:1976})
\begin{equation}\label{eq:Burnashev}
  E_B(R,Q) = B_Q \left( 1 - \nfrac{R}{C_Q} \right), 
\end{equation}
where 
\begin{align} \label{eq:B-term}
  B_Q &= \max_{x_A, x_R \in \ALPHABET X} b_Q (x_A, x_R), \\
  b_Q(x_A, x_R) &= 
  D\big( Q(\cdot | x_A) \| Q(\cdot | x_B) \big),
\end{align}
$Q(\cdot | x)$ is the probability distribution of the channel output when the
channel input is $x$, and
\begin{equation*}
  D(p \| q) = \sum_{y \in \ALPHABET Y} p(y) \log \frac {p(y)}{q(y)} 
\end{equation*}
is the Kullback-Leibler divergence between probability distributions $p$ and
$q$. We call $E_B(R,Q)$ as the \emph{Burnashev exponent} of channel $Q$ at rate
$R$ and $B_Q$ as the \emph{zero rate Burnashev exponent}. 

One of the key features of the Burnashev exponent is that it has a non-zero slope at
capacity. This slope captures the main advantage of feedback---by
reducing the transmission rate by a small \emph{fraction} of the capacity, we
linearly increase the error exponent, and therefore, exponentially decrease
the probability of error. Does feedback provide the same advantage for a
compound channel? 

Clearly, a particular component $E_\ell$ of the EER of the compound channel
cannot beat the Burnashev exponent for DMC $Q_\ell$. Thus, a trivial upper bound
for the EER at rate $\VEC R \in \ALPHABET C_{OF}(\ALPHABET Q)$ is the
hyper-rectangle with upper corner
\begin{equation}\label{eq:trivial}
  \left(
  B_{Q_1} \left(1 - \nfrac {R_1}{C_{Q_1}} \right), \dots,
  B_{Q_L} \left(1 - \nfrac {R_L}{C_{Q_L}} \right) \right)
\end{equation}
Tchamkerten and
Telatar~\cite{TchamkertenTelatar:2006a} showed that this bound is not tight
by means of a simple counterexample. They considered a compound family consisting
of two binary symmetric channels with complementary cross-over probabilities,
$p$ and $(1-p)$, where $p$ is known to the transmitter and the receiver. They showed
that, even for this simple family, no coding scheme universally achieves the
Burnashev exponent. 

Another way to interpret that result is that the EER need not be a
hyper-rectangle \emph{i.e.}, for a fixed rate $\SET R = \VEC R$ if $\VEC {E'},
\VEC {E''} \in \ALPHABET E(\SET R)$, then it is \emph{not necessary} that
\[ 
\left( \max( E'_1, E''_1), \dots, \max(E'_L, E''_L) \right) \in \ALPHABET E(\SET
R).\]
Thus, different sequence of coding schemes that achieve
the same rate vector $\VEC {R}$ may have different and \emph{non-comparable} error
exponents. Thus, in terms of error exponents, the compound channel with
feedback behaves in a manner similar to multi-terminal communication
channels~\cite{WengPradhanAnastasopoulos:2008}.

Tchamkerten and Telatar~\cite{TchamkertenTelatar:2006a} also identified
necessary and sufficient conditions on the compound family $\ALPHABET Q$ under
which the upper bound of~\eqref{eq:trivial} is tight for all rates along the
principle diagonal $(\gamma C_{Q_1}, \dots, \gamma C_{Q_L})$, $0 \le \gamma <
1$, of the opportunistic capacity region. For channels that do not satisfy these
conditions, the EER is not characterized. Even when these conditions are
satisfied, the EER is not characterized for rate vectors that are off the
principle diagonal (\emph{i.e.} $R_\ell/C_{Q_\ell}$ is not constant for all
$\ell = 1,\dots,L$). In Section~\ref{sec:scheme}, we present a coding scheme
for all rates in the opportunistic capacity region. This scheme 
achieves an \emph{error exponent with a non-zero slope} at all points in the
rate region, including points near the capacity boundary. This shows that
feedback provides similar advantage for a compound channel as for a DMC.

%
%

\subsection*{Notation} 

We use the following notation in this paper. $\PSP(\ALPHABET X)$ denotes the space of
probability distributions over $\ALPHABET X$. $\naturalnumbers$ denotes the set
of natural numbers. $\PR(\cdot)$ denotes the probability of an event,
$\EXP[\cdot]$ denotes the expectation of a random variable, and $\IND\{\cdot\}$
denotes the indicator function. All logarithms are to the base~$2$, and
$\exp_2(\cdot)$ denotes $2^{(\cdot)}$.

$C_Q$ denotes the capacity of the of a DMC with transition matrix
$Q$; $B_Q$ denotes its zero-rate Burnashev exponent. Given a compound family $\ALPHABET Q
= \SEQ Q$, $Q_\circ$ denotes the realized channel; $C_\ell$ denotes
the capacity $C_{Q_\ell}$ of DMC $Q_\ell$; $B_\ell$ denotes the zero-rate Burnashev
exponent of DMC $Q_\ell$. $\PR_\ell(\cdot)$ is short hand for
$\PR(\cdot | Q_\circ = Q_\ell)$; and $\EXP_\ell[\cdot]$ is a short hand for
$\EXP_\ell[\cdot | Q_\circ = Q_\ell]$. 

\section {Opportunistic capacity and error exponents} \label{sec:model}

In this section we formally define opportunistic capacity and error exponent regions for a
compound channel with feedback. Conceptually, it is easier to first define
achievable rate vector for fixed length communication and then extend that
definition to variable length communication. However, for succinctness, we only
define achievable rate vector for variable length communication. 

\begin{definition}[Variable-rate variable-length coding scheme]
A \emph{variable-rate variable-length coding scheme} for communicating over a compound channel
$\ALPHABET Q = \SEQ Q$ with feedback is a tuple $(\SET M, \SET f,
\SET g, \tau)$ where
\begin{itemize}
  \item $\SET M = \VEC M$ is the \emph{compound message size} 
    where $M_\ell \in \naturalnumbers$, $\ell = 1,\dots, L$. 
    Define $\ALPHABET M = \prod_{\ell=1}^L \{1, \dots, M_\ell \}$.

  \item $\SET f = (f_1, f_2, \dots)$ is the \emph{encoding strategy} where 
    \[ f_t : \ALPHABET M  \times \ALPHABET Y^{t-1} \mapsto
    \ALPHABET X, \quad t \in \naturalnumbers \]
    is the encoding function used at time $t$.
  \item $\SET g = (g_1, g_2, \dots)$ is the \emph{decoding strategy} where
    \[ g_t : \ALPHABET Y^t \mapsto \bigcup_{\ell=1}^L \{ (\ell,1), (\ell,2), \dots, (\ell,
    M_\ell) \}, \quad t \in
    \naturalnumbers \]
    is the decoding function at time $t$. 

  \item $\tau$ is the stopping time with respect to the channel outputs $Y^t$.
    More precisely, $\tau$ is a stopping time with respect to the filtration
    $\{2^{\ALPHABET Y^t}, t \in \naturalnumbers \}$. 
\end{itemize}
\end{definition}

The coding scheme is known to both the transmitter and the receiver. Variable
length communication takes place as follows. A \emph{compound message} $\SET W =
\VEC W$ is generated such that $W_\ell$ is uniformly distributed in $\{1,
\dots, M_\ell\}$.\footnote{All the probabilities of interest only depend on the
marginal distributions of $W_1$, \dots, $W_L$. So, the joint distribution of
$\VEC W$ need not be specified. \gobble{We explain the joint
for a particular application in Section~\ref{sec:remarks}.}} The transmitter uses
the encoding strategy $(f_1, f_2, \dots)$ to generate channel inputs 
\begin{equation*}
  X_1 = f_1(\SET W), \quad
  X_2 = f_2(\SET W, Y_1), \quad
  \cdots
\end{equation*}
until the stopping time $\tau$ with respect to the channel outputs. ($\tau$ is
known to the transmitter because of feedback.) The decoder then generates a
decoding decision
\[(\hat W, \hat L) = g_\tau(Y_1, \dots, Y_\tau).\]
The decoding decision consists of two components: the index $\hat L$ of 
decoded component and an estimate $\hat W$ of the $\hat L$-component
of the compound message $\SET W$. A communication error occurs if $\hat W \neq
W_{\hat L}$. 

\begin{remark} \label{rem:1}
  The above scheme is a variable-rate variable-length coding scheme. The
  transmitter and receiver agree upon the set of rates $\SEQ R$ before the
  start of communication. The transmitter chooses $L$ different messages, one
  message for each rate; At the end of communication, the receiver decides the
  message $\hat L$ it wants to decode and generates an estimate $\hat W$ for that
  message. Because of noiseless feedback, the encoder knows what the decoder
  decoded. In principle, the index $\hat L$ need not be the same as the index $\ell$ of the
  realized channel. For that reason, $\{\hat L \neq \ell\}$ is not considered a
  communication error.
\end{remark}

The two main performance metrics of a coding scheme are its error probability
and rate, both of which are vectors (rather than scalars), and denoted by $\SET
P = \VEC P$ and $\SET R = \VEC R$, respectively.
These are defined as follows.


\begin{definition}[Probability of error]
  A communication error occurs when $\hat W \neq W_{\hat L}$. The probability of
  error $\SET P = \VEC P$ of a coding scheme $(\SET M, \SET f, \SET g, \tau)$ is
  given by
  \[ P_\ell = \PR_\ell( \hat W \neq W_{\hat L}) \]
  where $\PR_\ell(\cdot)$ is a short hand notation for $\PR( \cdot | Q_{\circ} =
  Q_\ell )$. 
\end{definition}

\begin{definition}[Rate]
  The rate $\SET R = \VEC R$ of  a coding scheme $(\SET M, \SET f,
  \SET g, \tau)$ is given by
  \[ R_\ell = \frac{ \EXP_\ell[\log M_{\hat L}] }{\EXP_\ell[\tau]} \] 
  where $\EXP_\ell[\cdot]$ is a short hand notation for $\EXP[\cdot | Q_{\circ}
  = Q_\ell]$. 
\end{definition}

\begin{remark}
  The above scheme is a variable rate
  communication scheme. The size $M_{\hat L}$ of the communicated message
  $W_{\hat L}$ is a random variable taking values in $\SEQ M$. For that reason,
  we define the rate as $\EXP_\ell[ \log M_{\hat L}]/\EXP_\ell[\tau]$. When all
  rates $\SEQ R$ are equal, the above scheme reduces to a fixed-rate
  variable-length coding scheme and the definition of rate in
  Definition~\ref{def:rate} collapses to the
  traditional definition of fixed-rate variable-length coding.
\end{remark}

Rate and probability of error give rise to two asymptotic performance metrics,
\emph{viz.}, opportunistically achievable rate and error exponents. These are
defined as follows.

\begin{definition}[Opportunistically achievable rate]
  \label{def:rate}
A  rate vector $\SET R = \VEC R$ is said to be
\emph{opportunistically achievable} if
there exists a sequence of variable-rate variable-length coding schemes $(\SET M^{(n)}, \SET
f^{(n)}, \SET g^{(n)}, \tau^{(n)})$, $n \in \naturalnumbers$ such that:
\begin{enumerate}
  \item $\lim_{n \to \infty} \EXP_\ell[\tau^{(n)}] = \infty$ for $\ell =
    1,\dots, L$.
  \item For every $\varepsilon > 0$, there exists a $n_\circ(\varepsilon)$ so that
    for every $n \ge n_\circ(\varepsilon)$, we have
    \[ P_\ell^{(n)} < \varepsilon \quad \text{and} \quad R_\ell^{(n)} >
    R_\ell - \varepsilon, \quad \text{for all } \ell = 1,\dots, L; \]
    or equivalently,
    \[ \lim_{n \to \infty} P_\ell^{(n)} = 0 \quad \text{and} \quad
       \lim_{n \to \infty} R_\ell^{(n)} = R_\ell.
    \]
\end{enumerate}
\end{definition}

\begin{definition}[Opportunistic Capacity]
The union of all opportunistically achievable rates is called the \emph{opportunistic capacity
region} of the compound channel $\ALPHABET Q$ with feedback and denoted by $\ALPHABET
C_{OF}(\ALPHABET Q)$. 
\end{definition}
In Corollary~1, we show that $\ALPHABET C_{OF}(\ALPHABET Q)$ is given by
a hyper-rectangle with upper corner $(C_{Q_1}, \dots, C_{Q_L})$. For that
reason, we call $\ALPHABET C_{\ALPHABET Q} \DEFINED (C_{Q_1}, \dots, C_{Q_L})$
as the \emph{capacity vector} of the compound channel~$\ALPHABET Q$.

The variable-rate variable-length coding scheme defined above  is related to the
notion of rateless codes used in fountain
codes~\cite{ByersLubyMitzenmacherRege:1998, Luby:2002, Shokrollahi:2006} for
BER (binary erasure channel).

\begin{definition}[Error exponent]
Given a sequence of coding schemes $(\SET M^{(n)}, \SET f^{(n)}, \SET g^{(n)},
\tau^{(n)})$, $n \in \naturalnumbers$, that achieve a rate vector $\SET R$, the
asymptotic exponent $E_\ell$ of error probability $P_\ell$ is given by
\[E_\ell = \lim_{n \to \infty} - \frac{ \log P_\ell^{(n)}}{\EXP_\ell[\tau^{(n)}]}.\]
Then $\SET E = \VEC E$ is the error exponent of the sequence of
coding schemes $(\SET M^{(n)}, \SET f^{(n)}, \SET g^{(n)}, \tau^{(n)})$, $n \in \naturalnumbers$.
\end{definition}

\begin{definition}[Error exponent region]
  For a particular rate $\SET R$, the union of all possible error exponents is
  called the \emph{the error exponent region} (EER) of a compound channel with
  feedback and denoted by $\ALPHABET E(\SET R)$.
\end{definition}

In this paper, we study the EER for all rates in the
opportunistic capacity region and present lower bounds on the EER. 


The above scheme describes a variable-rate variable-length coding scheme;
varying the rate of the coding scheme allows for an additional degree of
freedom. This additional freedom does not affect the
opportunistic capacity region of compound channel; all rates within $\ALPHABET
C_{OF}(\ALPHABET Q)$ defined above can be achieved using a fixed-rate variable
length coding scheme. We do not know if this additional degree of freedom
improves the EER  since the EER of a compound channel has not been investigated
using the traditional fixed-rate variable-length coding scheme. The reason that we chose a
variable-rate coding scheme is that this additional degree of freedom
significantly simplifies the coding scheme.

\subsection* {Operational interpretation}

A transmitter has to reliably communicate an infinite bit stream, which is 
generated by a higher-layer application, to a receiver over a compound channel
with feedback. The transmitter uses a variable-rate variable-length 
coding scheme $(\SET M, \SET f, \SET g, \tau)$. For ease of
exposition, assume that every $M_\ell$, $\ell = 1,\dots, L$, is a power of $2$ so
that $\log M_\ell$ is an integer. Let $M^* = \max\SEQ M$ and $M_*
= \min\SEQ M$. The transmitter picks $\log M^*$ bits from the bit
stream. The decimal expansion of the first $\log M_\ell$ of these bits
determine the component $W_\ell$ of $\SET W$. The message $\SET W$ is
transmitted as described above. At stopping time $\tau$ the receiver passes
$(\hat L, \hat W)$ to a higher-layer application (which then converts $\hat W$
to bits) and the transmitter removes the first $\log M_{\hat L}$ bits%
from the
$\log M^*$ initially chosen bits and return the remaining $\log M^* - \log
M_{\hat L}$ bits to the bit stream. Then, the above process is repeated.

If the traditional pessimistic approach is followed, only $\log M_*$ bits are
removed from the bit stream at each stage. By following the opportunistic
approach, with high probability $\log M_\ell$ bits are removed from the bit
stream when the realized channel $Q_\circ$ is $Q_\ell$. By definition, $M_\ell \ge M_*$.
Thus, by defining capacity in an opportunistic manner, an additional $\log
M_\ell - \log M_*$ bits are removed at each step.

\subsection* {A trivial outer bound on error exponents}

Any coding scheme $(\SET M, \SET f, \SET g, \tau)$ for communicating over a
compound channel $\ALPHABET Q$ can also be used to communicate over DMC
$Q_\ell$. Hence, we have the following trivial upper bound on the EER.

\begin{proposition}
  For any variable-rate variable-length coding scheme for
  communicating over $\ALPHABET Q$ at rate $\VEC R$, each component
  of the error exponent region is bounded by the Burnashev exponent of channel
  $Q_\ell$, \emph{i.e.},
  \begin{equation*}
    E_\ell \le B_{Q_\ell}(1 - R_\ell/C_{Q_\ell})
  \end{equation*}
\end{proposition}

In the remainder of the paper, we try to derive a reasonable lower bound on the
EER.

\section {The coding scheme}
\label{sec:scheme}

In this section, we define a family of variable-rate variable-length coding schemes indexed by
$n \in \naturalnumbers$. As $n \to \infty$, the scheme opportunistically achieves a rate vector
$\VEC R$. This coding scheme is based on the
Yamamoto-Itoh~\cite{YamamotoItoh:1979} scheme that achieves the Burnashev
exponent for DMC. 

\subsection {Parameters of the coding scheme}

For each~$n \in \naturalnumbers$, the scheme is parameterized by
the following non-negative real constants:\footnote{The subscripts stand for
\emph{message} and \emph{control}.}
\begin{equation*}
  \alpha_m^{(n)}, \alpha_c^{(n)}, \text{ and } \beta_{m,\ell}^{(n)}, \beta_{c,
  \ell}^{(n)}, \xi_\ell^{(n)},\quad \ell = 1,\dots, L.
\end{equation*}
We will explain the purpose and choice of these constants later. For now, we
assume that $\alpha_m^{(n)}$, $\alpha_c^{(n})$, $\beta_{m,\ell}^{(n)}$ and
$\beta_{c,\ell}^{(n)}$ are chosen such that $\alpha_m^{(n)} n$, $\alpha_c^{(n)}
n$, $\beta_{m,\ell}^{(n)} n$ and $\beta_{c,\ell}^{(n)} n$ are integers. 
When there is no ambiguity, we will not explicitly show the dependence on $n$
and drop the superscripts~$^{(n)}$.


For each~$n$, the encoder and the decoder agree upon the following:
\begin{enumerate}
  \item \important{Two training sequences}, $\sigma_m$ and $\sigma_c$ of
    lengths $\alpha_m  n$ and $\alpha_c  n$ and \important{corresponding channel
    estimation rules} $\hat \theta_m$ and $\hat \theta_c$. 

  \item \important{$L$ codebooks}; one for each $Q_\ell$, $\ell = 1, \dots,
    L$. Codebook~$\ell$ has rate $\xi_\ell R_\ell/ \beta_{m,\ell}$ and length
    $\beta_{m,\ell}  n$.

  \item \important{$2L$ control sequences}; two for each $Q_\ell$, $\ell =
    1,\dots,L$, \emph{viz.}\footnote{The subscripts stand for \emph{accept} and
    \emph{reject}.}  $\sigma_{A, \ell}$ and $\sigma_{R, \ell}$, both of length
    $\beta_{c,\ell}  n$ and \important{corresponding hypothesis testing rules}
    $\hat \theta_{H, \ell}$ for disambiguating $\sigma_{A, \ell}$ and
    $\sigma_{R, \ell}$ over DMC~$Q_\ell$.
\end{enumerate}
A compound message $\SET W^{(n)}$ is chosen at random such that component 
$W^{(n)}_\ell$, $\ell = 1,\dots, L$, is uniformly distributed over $\{1,
\dots, \exp_2( n \xi_\ell R_\ell) \}$.\footnote{The joint distribution of $\VEC
{W^{(n)}}$ does not matter.}

\subsection{Operation of the coding scheme}

The coding scheme transmits in multiple epochs indexed by~$k \in
\naturalnumbers$. Each epoch consists of four phases:
\begin{enumerate}
  \item \important{A fixed length training phase} of length $\alpha_m  n$.
    During this phase the transmitter sends the training sequence $\sigma_m$; both
    the transmitter and the receiver use the estimation rule $\hat \theta_{m}$
    to determine a channel estimate $\hat L_m = \hat L_m(k)$. 

  \item \important{A variable length message phase} of length $\beta_{m, \hat
    L_m} n$. The transmitter and receiver use codebook $\hat L_m$ to send
    component~$\hat L_m$ of the compound message $\SET W^{(n)}$. Let $W_m(k)$ denote
    the transmitted message and $\hat W_m(k)$ the decoded message. 

  \item \important{A fixed length re-training phase} of length $\alpha_c  n$.
    During this phase the transmitter sends the training sequence $\sigma_c$; both
    the transmitter and the receiver use the estimation rule $\hat \theta_{c}$
    to determine a channel estimate $\hat L_c = \hat L_c(k)$. 

  \item \important{A variable length control phase} of length $\beta_{c,\hat
    L_c}n$. If $W_m(k) = \hat W_m(k)$, the transmitter sends a control message
    $W_c(k) = \sigma_{A, \hat L_c}$; otherwise it sends $W_c(k) = \sigma_{R, \hat L_c}$.
    The receiver decodes the \emph{control message} using $\hat \theta_{H, \hat
    L_c}$.  Let $\hat W_c(k)$ denote the estimated control message.
\end{enumerate}
If $\hat W_c(k) = \sigma_{A, \hat L_c}$, then transmission stops and the
receiver declares $(\hat L_m(k), \hat W_m(k))$ as its final decision; otherwise,
the compound message is retransmitted in the next epoch.  Let $K^{(n)}$ denote the
epoch when communication stops, \emph{i.e.},
\[ K^{(n)} = \inf \{ k \in \naturalnumbers : 
      \hat W_c(k) = \sigma_{A, \hat L_c(k)} \}. \]
Let the length of epoch~$k$ be $\Lambda^{(n)}(k) n$, \emph{i.e.},
\[ \Lambda^{(n)}(k) n = \alpha_m^{(n)} n + \beta_{m, \hat L_m(k)}^{(n)} +
                        \alpha_c^{(n)} n + \beta_{c, \hat L_c(k)}^{(n)}. \]
Hence, the length of communication is
\[ \tau^{(n)} = \sum_{k = 1}^{K^{(n)}} \Lambda^{(n)}(k) n \] 

\subsection {Choice of training sequences}

As described earlier, the transmitter and the receiver agree upon two training
sequences, $\sigma_m$ and $\sigma_c$, of lengths $\alpha_m n$ and $\alpha_c n$,
respectively. The optimal choice of such training sequences falls under the
domain of experiment design for estimating unknown parameters. We assume that we
can find good training sequences for $\ALPHABET Q$; if not, we choose a simple
training sequence that cycles through all the channel inputs one-by-one. 

The transmitter and the receiver also agree upon two estimating rules, $\hat
\theta_m$ and $\hat \theta_c$. For a training sequence $\sigma$ of size $n$ and a
estimation rule $\hat \theta$, define the estimation error exponent as
\begin{align}
  T^{\ell,k} &= \lim_{n \to \infty} {} - \frac 1n
  \log \PR_\ell( \hat \theta(Y^n) = k \mid X^n = \sigma), \quad k,\ell = 1,\dots,L \\
  \shortintertext{and for $\ell = 1,\dots, L$,}
  T_\ell &= \lim_{n \to \infty} - \frac 1n 
  \log \PR_\ell( \hat \theta(Y^n) \neq \ell \mid  X^n = \sigma ) \notag \\
  &= \min \{T^{\ell,k} : k = 1, \dots, L, k \neq \ell\} \label{eq:relation}
\end{align}
where $X^n$ and $Y^n$ are the channel inputs and outputs respectively. We are
interested in characterizing the union of $\VEC T$ for all choices of estimation
rule $\hat \theta$. We call this region the \emph{estimation error exponent
region} and denote it by $\ALPHABET T$. Instead of directly characterizing
estimation error exponent region, it is easier to first characterize
\emph{pairwise estimation error exponent region}---the union of $(T^{\ell,k}; \ell, k
= 1, \dots, T$; $k \neq \ell)$ for all choices of estimation rule $\hat \theta$;
this region is denoted by $\ALPHABET T^*$---and then obtain the
estimation error estimation region $\ALPHABET T$ using~\eqref{eq:relation}.

Characterizing the pairwise estimation error exponent is equivalent to
characterizing the pairwise hypothesis testing exponent for multiple hypothesis
testing. The latter was characterized by Tuncel~\cite{Tuncel:2005} 
for $L$-ary hypothesis testing with independent and identically distributed
observations. Let $p_\ell$ be the probability distribution of the
observations under hypothesis $\ell$. Then,
\begin{equation*}
\ALPHABET T^* = \{ (T^{\ell,k}, \ell \neq k) : \forall p \in \PSP(Y) , 
\exists k 
\text{ such that } D(p \| p_\ell) \ge T^{\ell,k} \text{ for all } \ell \neq k\} 
\end{equation*}
For our setup, the observations at the receiver need not be identically distributed.
Nonetheless, the observations are independent across time, and it is easy to
generalize the above region to the case of independent (but not identically
distributed) observations. We then use~\eqref{eq:relation} to obtain the
desired region $\ALPHABET T$ as follows:
\begin{equation*}
  \ALPHABET T = \{ \VEC T : \exists (T^{\ell,k}, \ell \neq k) \in
  \ALPHABET T^*
  \text{ such that } \forall \ell, T_\ell = \min_{k \neq \ell} T_{\ell k} \}
\end{equation*}

The estimation rules $\hat \theta_m$ and $\hat \theta_c$ attain particular
points in $\ALPHABET T$; denote these by $(T_{m, 1}, \dots, T_{m,L})$ and $(T_{c,1},
\dots, T_{c,L})$, respectively. Recall that the training sequences $\sigma_m$
and $\sigma_c$ are of length $\alpha_m n$ and $\alpha_c n$ respectively. Thus, for any epoch $k$, 
\begin{gather}
  \lim_{n \to \infty} - \frac 1{\alpha_m n} \log \PR_\ell(\hat L_{m} \neq \ell) =
  T_{m,\ell}, \quad \ell = 1,\dots, L;
  \shortintertext{and}
  \lim_{n \to \infty} - \frac 1{\alpha_c n} \log \PR_\ell(\hat L_{c} \neq \ell) =
  T_{c,\ell}, \quad \ell = 1,\dots, L.
\end{gather}

Choose $\hat \theta_m$ and $\hat \theta_c$ such that
\begin{gather}
  \lim_{n \to \infty} \PR_\ell(\hat L_m \neq \ell) = 0, \quad
  \lim_{n \to \infty} \PR_\ell(\hat L_c \neq \ell) = 0; \\
  \shortintertext{and}
  T_{m, \ell} \ge 0, \quad T_{c,\ell} > 0, \quad \ell = 1,\dots, L
\end{gather}

\subsection {Choice of codebooks}

As described earlier, the transmitter and receiver agree upon $L$ codebooks.
Codebook $\ell$ is a fixed length codebook for DMC~$Q_\ell$, $\ell = 1, \dots,
L$, with rate $\xi_\ell R_\ell/ \beta_{m.\ell}$ and length $\beta_{m, \ell} n$.
Choose codebook~$\ell$ such that the error
exponent is positive for all rates below capacity, i.e.,
\begin{equation}
  \text{if } \frac {\xi_\ell R_\ell}{\beta_{m,\ell}} < C_\ell, 
  \text{ then } 
  \lim_{n \to \infty} - \frac {1}{\beta_{m,\ell} n} \log \PR_\ell( W_m(k) \neq \hat W_m(k) ) > 0
\end{equation}
The actual form of the codebook does not matter; for example, it could be a linear
code, or a convolutional code, or a LDPC code, or a polar code,  or a posterior
matching code that uses feedback.

\subsection {Choice of control sequences} \label{sec:control}

As described earlier, the transmitter and the receiver agree upon two control
sequences, $\sigma_{A,\ell}$ and $\sigma_{R,\ell}$ of length $\beta_{c,\ell}$, for
signaling \textsc{accept} (when $W_m = \hat W_m$) and \textsc{reject} (when $W_m
\neq \hat W_m$). Choose these sequences as
repetitions of $x_{A,\ell}$ and $x_{R,\ell}$, the maximally separated input
symbols for $Q_\ell$, \emph{i.e.}, the $\arg \max$ in~\eqref{eq:B-term} for
$B_{Q_\ell}$. 

The transmitter and the receiver also agree upon a hypothesis testing rule
$\hat \theta_{H,\ell}$ for disambiguating $\sigma_{A,\ell}$ and $\sigma_{R,\ell}$. Let
$H_{A,\ell}$ and $H_{R,\ell}$ denote the error exponents of this rule, that is,
\begin{gather}
  H_{A,\ell} = \lim_{n \to \infty} - \frac 1{\beta_{c,\ell} n} \log 
  \PR_\ell(\hat W_c(k) \neq W_c(k) \mid  W_c(k) = \sigma_{A,\ell});
  \shortintertext{and}
  H_{R,\ell} = \lim_{n \to \infty} - \frac 1{\beta_{c,\ell} n} \log 
  \PR_\ell(\hat W_c(k) \neq W_c(k) \mid  W_c(k) = \sigma_{R,\ell}).
\end{gather}
Choose $\hat \theta_{H,\ell}$ such that
\begin{equation}
  H_{A,\ell} = 0 \quad \text{and} \quad H_{R,\ell} = B_\ell;
\end{equation}
while
\begin{equation}
  \lim_{n \to \infty} \PR_\ell(\hat W_c(k) \neq W_c(k) \mid  W_c(k) = \sigma_{A,\ell}) = 
  \lim_{n \to \infty} \PR_\ell(\hat W_c(k) \neq W_c(k) \mid  W_c(k) = \sigma_{R,\ell}) = 
  0.
\end{equation}
Such a choice of $\hat \theta_{H,\ell}$ is always possible
(see~\cite{Blahut:1974}).

\subsection {Choice of parameters}

The first and second phase of the proposed scheme correspond to the
\emph{message mode} of the Yamamoto Itoh~\cite{YamamotoItoh:1979} scheme, while
the third and fourth phase correspond to the \emph{control mode}. In the
Yamamoto Itoh scheme, the ratio of the lengths of the message and control modes
is $\gamma/(1-\gamma)$ where $\gamma = R/C$. We choose the parameters such that
a similar relation holds for the proposed scheme. In particular, let
$\gamma_\ell = R_\ell/C_\ell$; then, we want
\begin{equation*}
  \lim_{n \to \infty}
  \frac{ \alpha_m + \beta_{m,\ell} } 
       { \alpha_c + \beta_{c,\ell} }
   = \frac {\gamma_\ell} {1-\gamma_\ell}.
\end{equation*}
The parameter $\xi_\ell$ is the proportionality constant, that is,
\begin{equation*}
  \lim_{n \to \infty} \alpha_m + \beta_{m, \ell}  = \xi_\ell \gamma_\ell
  \quad \text{and} \quad
  \lim_{n \to \infty} \alpha_c + \beta_{c, \ell}  = \xi_\ell (1 -
  \gamma_\ell).
\end{equation*}
We let one of these proportionality constants to be one and call that channel
the \emph{reference channel} $Q_*$.

In the Burnashev exponent, the slope (\emph{i.e.}, the $B_Q$ term
in~\eqref{eq:Burnashev}) is determined by the ``signaling exponent'' in the
control mode. As will become apparent in the proof of
Proposition~\ref{prop:exponent}, to maximize the slope of our exponent, we need
to choose the parameters such that
\begin{equation*}
  \lim_{n \to \infty} {}- \frac 1n \log
  \PR_\ell(\hat W_c(1) \neq W_c(1) \mid W_c(1) = \sigma_{R, \ell}, \hat L_c(1) = \ell) 
  = \lim_{n \to \infty} {}- \frac 1n \log
  \PR_\ell(\hat L_c(1) \neq \ell).
\end{equation*}

We choose the parameters that satisfy the above properties as follows.
For $\ell = 1,\dots, L$, define constants
\begin{equation}
  \kappa_\ell = \frac{T_{c,\ell}}{B_\ell}, \quad
  \gamma_\ell = \frac{R_\ell}{C_\ell}, \quad
  \zeta_\ell = \frac{(1 - \gamma_\ell)}{(1 + \kappa_\ell)}.
\end{equation}
Let $\kappa_*$, $\gamma_*$, and $\zeta_*$ be the $\kappa$, $\gamma$ and $\zeta$
parameters corresponding to the reference channel $Q_*$. 
Then choose the parameters of the coding scheme as
follows:
\begin{enumerate}
  \item Choose $\xi_\ell = \zeta_*/\zeta_\ell$.

  \item Choose $\alpha_m^{(n)} > 0$ such that $\alpha_m^{(n)} n$ is an integer, $\lim_{n \to \infty}
    \alpha_m^{(n)} = 0$ while $\lim_{n \to \infty} \alpha_m ^{(n)} n = \infty$.
    An example for such a choice is $\alpha_m^{(n)}  = \lfloor n/\log n
    \rfloor / n$. 

  \item Choose $\beta_{m,\ell}^{(n)} > \xi_\ell \gamma_\ell$ such that
    $\beta_{m,\ell}^{(n)}n$ is an integer, $\lim_{n \to \infty} \beta_{m,\ell} = \xi_\ell \gamma_\ell$.

  \item Choose $\alpha_c^{(n)} > 0$ such that $\alpha_c^{(n)} n$ is an integer, $\lim_{n \to \infty}
    \alpha_c^{(n)} = \zeta_*$.

  \item Choose $\beta_{c,\ell}^{(n)} > 0$ such that $\beta_{c,\ell}^{(n)}$ is an
    integer, $\lim_{n \to \infty}
    \beta_{c,\ell}^{(n)} = \kappa_\ell\zeta_*$.
\end{enumerate}

\subsection {Consequences of the choice of parameters}

The choice of the parameters $\alpha_m$, $\alpha_c$, $\beta_{m,\ell}$,
$\beta_{c,\ell}$, and $\xi_\ell$, $\ell = 1, \dots, L$ implies the following:

\begin{lemma}[Length of message and control phases]
  \label{lemma:length}
  For every $\ell = 1,\dots,L$, we have that
  \begin{equation*}
    \lim_{n \to \infty} \alpha_m + \beta_{m, \ell}  = \xi_\ell \gamma_\ell
    \quad \text{and} \quad
    \lim_{n \to \infty} \alpha_c + \beta_{c, \ell}  = \xi_\ell (1 -
    \gamma_\ell).
  \end{equation*}
\end{lemma}

The choice of the estimation rules $\hat \theta_m$, $\hat \theta_c$, the
codebooks, and the hypothesis testing rules $\hat \theta_{H,\ell}$, $\ell = 1,
\dots, L$, implies the following properties:

\begin{lemma}
  \label{lemma:exponent}
  For every $k \in \naturalnumbers$ and $\ell = 1,\dots,L$, we have that
  \begin{align*}
    \lim_{n \to \infty} & - \frac 1 {\alpha_m n} \log 
    \PR_\ell( \hat L_m(k) \neq \ell \mid K \ge k) = T_{m,\ell}; \\ 
    \lim_{n \to \infty} & - \frac 1 {\alpha_c n} \log 
    \PR_\ell( \hat L_c(k) \neq \ell \mid K \ge k) = T_{c,\ell}; \\
    \lim_{n \to \infty} & - \frac 1 {\beta_{m, \ell} n} \log 
    \PR_\ell( \hat W_m(k) \neq W_m(k) \mid \hat L_m(k) = \ell, K \ge k) > 0; \\
    \lim_{n \to \infty} & - \frac 1 {\beta_{c, \ell} n} \log 
    \PR_\ell( \hat W_c(k) \neq W_c(k) \mid \hat L_c(k) = \ell, W_c(k) =
    \sigma_{A,\ell}, K \ge k) = H_{A,\ell} = 0; \\
    \lim_{n \to \infty} & - \frac 1 {\beta_{c, \ell} n} \log 
    \PR_\ell( \hat W_c(k) \neq W_c(k) \mid \hat L_c(k) = \ell, W_c(k) =
    \sigma_{R,\ell}, K \ge k) = H_{R,\ell} = B_\ell. 
  \end{align*}
\end{lemma}

An immediate consequence of the above is that each of the error probabilities
approach zero as $n \to \infty$. Specifically,

\begin{lemma}
  \label{lemma:estimation}
  For every $k \in \naturalnumbers$ and $\ell = 1,\dots,L$, we have that
  \begin{align*}
    \lim_{n \to \infty} & 
    \PR_\ell( \hat L_m(k) \neq \ell \mid K \ge k) = 0; \\ 
    \lim_{n \to \infty} & 
    \PR_\ell( \hat L_c(k) \neq \ell \mid K \ge k) = 0; \\
    \lim_{n \to \infty} & 
    \PR_\ell( \hat W_m(k) \neq W_m(k) \mid \hat L_m(k) = \ell, K \ge k) = 0; \\
    \lim_{n \to \infty} &
    \PR_\ell( \hat W_c(k) \neq W_c(k) \mid \hat L_c(k) = \ell, W_c(k) =
    \sigma_{A,\ell}, K \ge k) = 0; \\
    \lim_{n \to \infty} &
    \PR_\ell( \hat W_c(k) \neq W_c(k) \mid \hat L_c(k) = \ell, W_c(k) =
    \sigma_{R,\ell}, K \ge k) = 0. 
  \end{align*}
\end{lemma}

\section {Performance analysis} \label{sec:performance}

\subsection {Some preliminary results} 

Recall that the length of epoch $k \in \naturalnumbers$ is $\Lambda(k) n$.
Thus, 
\[ \Lambda(k) = \alpha_m + \beta_{m, \hat L_m(k)} + \alpha_c + \beta_{c, \hat
L_c(k)}. \]
Combining Lemmas~\ref{lemma:length} and~\ref{lemma:estimation}, we get the
following:

\begin{lemma} \label{lemma:lambda}
  For every $k \in \naturalnumbers$ and $\ell = 1,\dots,L$, we have that
  \begin{equation}
    \lim_{n \to \infty} \EXP_\ell[\Lambda(k)] = \xi_\ell.
  \end{equation}
  Thus, for large $n$ and realized channel $Q_\ell$, the expected length of each epoch is $\xi_\ell n$.
\end{lemma}

Let 
\[ \Sigma_R = \{ \sigma_{R, \ell}, \ell = 1, \dots, L \} \]
denote the set of all \textsc{reject} control signals and 
let $\rho_\ell^{(n)}$ denote the probability that the estimated  control sequence 
in epoch~$k$ is in $\Sigma_R$, \emph{i.e.},
\begin{equation}
  \rho_\ell = \PR_\ell( \hat W_c(k) \in \Sigma_R
  \mid K \ge k)
\end{equation}
Due to symmetry across each epoch, $\rho_\ell$ does not depend on~$k$.

Conditioned on the event that $K \ge k$, communication stops at epoch $k$ if the
estimated control sequence $\hat W_c(k)$ is \textsc{reject}. Hence,
\begin{equation*}
  \PR_\ell( K = k \mid K \ge k) = \rho_\ell.
\end{equation*}
Consequently, we have the following:

\begin{proposition} \label{prop:K}
  For any $n \in \naturalnumbers$ and $\ell = 1,\dots,L$, the number
  of retransmissions has a geometric distribution; in particular,
  \begin{equation}
    \PR_\ell(K = k) = \rho_\ell (1 - \rho_\ell)^{k-1}, \quad k \in \naturalnumbers
  \end{equation}
  Furthermore, Lemma~\ref{lemma:estimation} implies that
  \begin{equation}
    \lim_{n \to \infty} \rho_\ell^{(n)} = 1.
  \end{equation}
  Hence,
  \begin{equation}
    \lim_{n \to \infty} \PR_\ell(K^{(n)} = 1) = 1.
  \end{equation}
  Thus, for large $n$ and irrespective of the realized channel, the expected
  number of transmission epochs is one.
\end{proposition}

\subsection {Expected length of communication} 

\begin{proposition} \label{prop:stopping-time}
  For every $\ell = 1,\dots,L$,
  \begin{equation}
    \lim_{n \to \infty} \frac 1n \EXP_\ell[ \tau^{(n)} ] = \xi_\ell
  \end{equation}
\end{proposition}
\begin{proof}
  Since $\tau = \sum_{k=1}^K \Lambda(k) n$, we get
  \begin{gather*}
    \frac 1n \EXP_\ell[\tau] = \EXP_\ell\bigg[ \sum_{k=1}^K \Lambda(k) \bigg]
    = \PR_\ell(K = 1) \EXP_\ell[\Lambda(1)] + \PR_\ell(K > 1) 
       \EXP_\ell\bigg[ \sum_{k=1}^K \Lambda(k) \biggm| K > 1 \bigg] \\
   \intertext{Proposition~\ref{prop:K} implies that}
   \lim_{n \to \infty} \frac 1n \EXP_\ell[\tau] = 
   \lim_{n \to \infty} \EXP_\ell[ \Lambda(1)] = \xi_\ell
  \end{gather*}
  where the last equality follows from Lemma~\ref{lemma:lambda}.
\end{proof}

\subsection {Probability of error}

\begin{proposition}
  \label{prop:perr}
  For any $n \in \naturalnumbers$ and $\ell = 1,\dots,L$, the probability of
  error is given by
  \begin{equation}
    P_\ell^{(n)} = \frac 1{\rho_\ell^{(n)}} 
    \PR_\ell(\hat W_m(1) \neq W_m(1))
    \PR_\ell(\hat W_c(1) \neq W_c(1) \mid W_c(1) \in \Sigma_R)
  \end{equation}
\end{proposition}
\begin{proof}
  The error event is $\{\hat W_m(K) \neq W_{\hat L_m(K)}\}$. For each $k \in
  \naturalnumbers$, $W_m(k) = W_{\hat L_m(k)}$. Using this to simplify the
  probability of error, we get that
  \begin{align*}
    P^{(n)}_\ell &= \PR_\ell( \hat W_m(K) \neq W_m(K) ) \\
    &= \sum_{k=1}^\infty \PR_\ell( \hat W_m(K) \neq W_m(K), K = k) \\
    &= \sum_{k=1}^\infty \PR_\ell( K = k) \PR_\ell( \hat W_m(k) \neq W_m(k) \mid K = k) \\
    &\stackrel{(a)}= 
    \PR_\ell( \hat W_m(1) \neq W_m(1) \mid K = 1) 
    \sum_{k=1}^\infty \PR_\ell( K = k) \\
    &\stackrel{(b)}= \frac 1{\PR_\ell(K = 1)} 
    \PR_\ell( \hat W_m(1) \neq W_m(1)) \PR_\ell(K = 1 \mid \hat W_m(1) \neq W_m(1))
    \\
    &= \frac 1{\PR_\ell(K = 1)} 
    \PR_\ell(\hat W_m(1) \neq W_m(1))
    \PR_\ell(\hat W_c(1) \neq W_c(1) \mid W_c(1) \in \Sigma_R)
  \end{align*}
  where $(a)$ follows from the symmetry across epochs and $(b)$ follows from
  Bayes rule.
\end{proof}

\subsection {Opportunistically achievable rate}

\begin{proposition}
  The coding scheme of Section~\ref{sec:scheme} opportunistically achieves the
  rate vector $\VEC R$.
\end{proposition}
\begin{proof}
  To prove the result, we need to show the proposed scheme satisfies the 
  properties described in Definition~\ref{def:rate}. Specifically,
  \begin{equation} \label{eq:tau:*}
   \lim_{n \to \infty} \EXP_\ell[ \tau ] = \infty;
  \end{equation}
  along with
  \begin{equation}
    \lim_{n \to \infty} 
    \frac{ \EXP_\ell[ \log M_{\hat L_m(K)} ]} { \EXP_\ell[\tau] } = R_\ell;
    \label{eq:rate:*}
  \end{equation}
  and
  \begin{equation}
    \lim_{n \to \infty} P_\ell^{(n)} = 0.
    \label{eq:perr:*}
  \end{equation}

  We prove these separately.
  \begin{enumerate}[(a)]
    \item Property~\eqref{eq:tau:*} follows from
      Proposition~\ref{prop:stopping-time}.

    \item Recall that $M_\ell = \exp_2(n \xi_\ell R_\ell)$. Hence,
      \begin{gather}
        \frac{ \EXP_\ell[ \log M_{\hat L_m(K)} ]} { \EXP_\ell[\tau] } = 
        \EXP_\ell[\xi_{\hat L_m(K)} R_{\hat L_m(K)}] 
        \frac{n}{\EXP_\ell[\tau]}; \notag
        \shortintertext{Proposition~\ref{prop:stopping-time} implies that}
        \lim_{n \to \infty} 
        \frac{ \EXP_\ell[ \log M_{\hat L_m(K)} ]} { \EXP_\ell[\tau] } = 
        \frac 1 {\xi_\ell} \lim_{n \to \infty} \EXP_\ell[ \xi_{\hat L_m(K)} R_{\hat L_m(K)} ],
        \label{eq:rate:1}
      \end{gather}
      Now, 
      \begin{gather}
        \EXP_\ell[ \xi_{\hat L_m(K)}  R_{\hat L_m(K)} ] = \PR_\ell(K = 1) \EXP_\ell[ \xi_{\hat L_m(1)}  R_{\hat L_m(1)} ]
        + \PR_\ell(K > 1) \EXP_\ell[  \xi_{\hat L_m(K)} R_{\hat L_m(K)} \mid K > 1] \notag \\
        \intertext{Using Proposition~\ref{prop:K} we get that}
        \lim_{n \to \infty} \EXP_\ell[ \xi_{\hat L_m(K)} R_{\hat L_m(K)} ] = \lim_{n \to \infty}
        \EXP_\ell[ \xi_{\hat L_m(1)}  R_{\hat L_m(1)} ] \label{eq:rate:2}
      \end{gather}
      Now,
      \begin{gather}
        \begin{multlined}
        \EXP_\ell[  \xi_{\hat L_m(1)} R_{\hat L_m(1)} ] = 
        \PR_\ell(\hat L_m(1) = \ell) \EXP_\ell[  \xi_{\hat L_m(1)} R_{\hat L_m(1)} \mid \hat L_m(1) = \ell] 
        \\
        +
        \PR_\ell(\hat L_m(1) \neq \ell) \EXP_\ell[  \xi_{\hat L_m(1)} R_{\hat L_m(1)} \mid \hat L_m(1) \neq \ell] 
      \end{multlined}
        \notag \\
        \intertext{Using Lemma~\ref{lemma:estimation} we get that}
        \lim_{n \to \infty} \EXP_\ell[\xi_{\hat L_m(1)} R_{\hat L_m(1)}] = \xi_\ell R_\ell
        \label{eq:rate:3}
      \end{gather}
      Substituting~\eqref{eq:rate:2} and~\eqref{eq:rate:3} in~\eqref{eq:rate:1}
      gives~\eqref{eq:rate:*}.

    \item Property~\eqref{eq:perr:*} follows substituting the results of
      Proposition~\ref{prop:K} and Lemma~\ref{lemma:estimation} in
      Proposition~\ref{prop:perr}.
  \end{enumerate}

\end{proof}

\subsection{Error exponent region}

\begin{proposition}
  \label{prop:exponent}
  For a particular choice of estimation rule $\hat \theta_c$, the
  $\ell$-component of the error exponent $\VEC E$ of the coding scheme of
  Section~\ref{sec:scheme} is bounded by
  \begin{equation}
    E_\ell \ge \frac{\kappa_\ell}{1+\kappa_\ell} B_\ell(1-\gamma_\ell)
    = \frac{T_{c,\ell}}{T_{c,\ell} + B_\ell} B_\ell \left( 1 - \frac
    {R_\ell}{C_\ell} \right).
  \end{equation}
  By varying the choice of $\hat \theta_c$, we get
  \begin{equation}
    \ALPHABET E \VEC R \supseteq 
    \bigcup_{(T_{c,1}, \dots, T_{c,L}) \in \ALPHABET T}
    \left(
    \frac{T_{c,1}}{T_{c,1} + B_1} B_1 \left( 1 - \frac {R_1}{C_1} \right),
    \dots,
    \frac{T_{c,L}}{T_{c,L} + B_L} B_L \left( 1 - \frac {R_L}{C_L} \right)
    \right)
  \end{equation}
\end{proposition}
\begin{proof}
  Consider the expression for $P_\ell$ in Proposition~\ref{prop:perr}. Taking
  logs, we get
  \begin{equation}
    \begin{split}
      {}- \frac 1n \log P_\ell &= 
      {}- \frac 1n \log \PR_\ell( \hat W_m(1) \neq W_m(1)) \\
      &\quad {} - \frac 1n \log \PR_\ell( \hat W_c(1) \neq W_c(1) \mid W_c(1)
      \in \Sigma_R) \\
      & \quad {} + \frac {\log \rho_\ell}{n}
    \end{split}
    \label{eq:log_P}
  \end{equation}

  Consider the three summands in the RHS of~\eqref{eq:log_P}. First consider the
  first term of of the RHS of~\eqref{eq:log_P}. From
  Lemma~\ref{lemma:estimation}, we have that
  \begin{equation}
    \lim_{n \to \infty} {} - \frac 1n \log \PR_\ell( \hat W_m(1) \neq W_m(1) ) >
    0.
    \label{eq:log_P:1}
  \end{equation}

  Next consider the second term of the RHS of~\eqref{eq:log_P}.
  \begin{multline}
    \PR_\ell(\hat W_c(1) \neq W_c(1) \mid W_c(1) \in \Sigma_R) 
    \\ < 
    \PR_\ell(\hat W_c(1) \neq W_c(1) \mid W_c(1) = \sigma_{R, \ell}, \hat L_c(1) = \ell) 
    +
    \PR_\ell(\hat L_c(1) \neq \ell) 
    \label{eq:log_P:2}
  \end{multline}
  From Lemma~\ref{lemma:estimation}, we have that
  \begin{equation}
    \lim_{n \to \infty} {}- \frac 1n \log
    \PR_\ell(\hat W_c(1) \neq W_c(1) \mid W_c(1) = \sigma_{R, \ell}, \hat L_c(1) = \ell) 
    = \kappa_\ell \zeta_* B_\ell ; \label{eq:log_P:3}
  \end{equation}
  and
  \begin{equation}
    \lim_{n \to \infty} {}- \frac 1n \log
    \PR_\ell(\hat L_c(1) \neq \ell) 
    = \zeta_* T_{c,\ell} = \zeta_* \kappa_\ell B_\ell. 
    \label{eq:log_P:4}
  \end{equation}
  where the last equality follows because $\kappa_\ell = T_{c,\ell}/B_\ell$. 
  Substituting~\eqref{eq:log_P:3} and~\eqref{eq:log_P:4} in~\eqref{eq:log_P:2},
  and taking logarithms and limits, we get
  \begin{equation}
    \lim_{n \to \infty} {}- \frac 1n \log
    \PR_\ell(\hat W_c(1) \neq W_c(1) \mid W_c(1) \in \Sigma_R) 
    > \zeta_* \kappa_\ell B_\ell. 
    \label{eq:log_P:5}
  \end{equation}
  
  Next consider the third term of the RHS of~\eqref{eq:log_P}. From
  Proposition~\ref{prop:K}, it follows that
  \begin{equation}
    \lim_{n \to \infty} {} \frac {\log \rho_\ell}n = 0.
    \label{eq:log_P:6}
  \end{equation}

  Substituting the result of~\eqref{eq:log_P:1}, \eqref{eq:log_P:5},
  and~\eqref{eq:log_P:6} in~\eqref{eq:log_P}, we get
  \begin{equation}
    {}- \frac 1n \log P_\ell > \zeta_* \kappa_\ell B_\ell.
  \end{equation}
  Combining this with Proposition~\ref{prop:stopping-time}, we get
  \begin{equation}
    E_\ell = {}- \frac 1{\EXP_\ell[\tau]} \log P_\ell > \zeta_* \kappa_\ell
    B_\ell/ \xi_\ell  \label{eq:log_P:8}
  \end{equation}
  The result follows by observing that $\xi_\ell = \zeta_*/\zeta_\ell$, and
  substituting the value of $\kappa_\ell$ and $\zeta_\ell$
  in~\eqref{eq:log_P:8}.
\end{proof}

The choice of operating point on the EER boundary depends on the objective.
For given positive constants $w_1$, \dots, $w_L$, two possible objectives are to
minimize the weighted probability of error
\[ \bar P \DEFINED (w_1 P_1 + \dots + w_L P_L)/(w_1 + \dots + w_L) \]
or maximize the weighted error exponent
\[ \bar E \DEFINED (w_1 E_1 + \dots + w_L E_L)/(w_1 + \dots + w_L) \]
As $n \to \infty$, each of
$P_1$, \dots, $P_L$ decay to zero exponentially. Thus, minimizing $\bar P$ is
equivalent to maximizing $\min \SEQ E$. The choice of the operating point $\VEC
E$, and hence the choice of $\hat \theta_c$, depends on the objective.

\subsection{Capacity}

Proposition~\ref{prop:exponent} implies that for any rate vector $\VEC R$ such
that $R_\ell < C_\ell$, $\ell = 1,\dots, L$, each component of the probability
of error $\VEC P$ goes to zero as $n \to \infty$. Thus,
\begin{equation*}
  \ALPHABET C_{OF}(\ALPHABET Q) \supseteq \big\{ \VEC R :
  0 \le R_\ell < C_{\ell}, \ \ell=1,\dots,L \big\}. 
\end{equation*}

Furthermore, if a coding scheme (opportunistically) achieves rate $R_\ell$ when
the realized channel $Q_\circ = Q_\ell$, then the same scheme will also achieve
rate $R_\ell$ when used over DMC $Q_\ell$. Thus,
\begin{equation*}
  \ALPHABET C_{OF}(\ALPHABET Q) \subseteq \big\{ \VEC R :
  0 \le R_\ell < C_{\ell}, \ \ell=1,\dots,L \big\}. 
\end{equation*}

Combining these two bounds, we get
\begin{corollary}\label{cor:capacity}
  The opportunistic capacity region is given by a
  hyper-rectangle
  \begin{equation*}
    \ALPHABET C_{OF}(\ALPHABET Q) = \big\{ \VEC R :
    0 \le R_\ell < C_{\ell}, \ \ell=1,\dots,L \big\}. 
  \end{equation*}
  We call $\ALPHABET C_{\ALPHABET Q} \DEFINED (C_1, \dots, C_L)$ as called
  the \emph{capacity vector} of the compound channel~$\ALPHABET Q$. 
\end{corollary}

\section {An example} \label{sec:example}

Consider a compound channel consisting of two BSCs with
complementary crossover probabilities, $p$ and $(1-p)$, where $0 < p < 1/2$ and
$p$ is known to the transmitter and the receiver. Denote this compound channel
by
\begin{equation*}
  \ALPHABET Q_p \DEFINED \{\BSC_p, \BSC_{1-p}\}
\end{equation*}
where $\BSC_p$ denotes a binary symmetric channel with crossover probability
$p$. For convenience, we index all variables by $p$ and $(1-p)$ rather than
by $1$ and $2$. For binary symmetric channel, the capacity and the zero-rate
Burnashev exponent are given by
\begin{gather*}
  C_p = C_{1-p} = 1 - h(p) \\
  \shortintertext{and}
  B_p = B_{1-p} = D(p \| 1-p)
\end{gather*}
where $h(p) = -p \log p - (1-p) \log (1-p)$ is the binary entropy function and
$D(p\|q) = p \log (p/q) + (1-p) \log ( (1-p)/(1-q))$ is the binary
Kullback-Leibler function. Assume that the desired communication rate is $(R_p,
R_{1-p})$, where $R_p < C_p$ and $R_{1-p} < C_{1-p}$.

Choose the training sequences $\sigma_m$ and $\sigma_c$ as all zero sequences of
length $\alpha_m n $ and $\alpha_c n$. Choose the channel estimation rules $\hat
\theta_m$ and $\hat \theta_c$ as the threshold tests: if the empirical frequency
of ones in the output is less than $q$, $p < q < 1-p$, estimate the channel as
$\BSC_p$; otherwise, estimate the channel as $\BSC_{1-p}$. The thresholds for
$\hat \theta_m$ and $\hat \theta_c$ are $q_m$ and $q_c$ respectively. For such a
threshold test, the probability of estimation error is bound by the tail 
probability of a sum of independent random variables. From Hoeffding's
inequality~\cite[Theorem~1]{Hoeffding:1963}, the exponents of the estimation
errors are
\begin{equation*}
  T_{m,p} = D(q_m \| p), \quad T_{m,1-p} = D(q_m \| 1-p), \quad
  T_{c,p} = D(q_c \| p), \quad T_{c,1-p} = D(q_c \| 1-p).
\end{equation*}

Choose the two codebooks as any codebooks for $\BSC_p$ and $\BSC_{1-p}$ that
have positive error exponents. 

Choose the control sequences $\sigma_{A,p}$ and $\sigma_{R,p}$ as $\beta_{c,p}n$
repetitions of zeros and ones, respectively. Similarly, choose the control
sequences $\sigma_{A,1-p}$ and $\sigma_{R,1-p}$ as $\beta_{c,1-p}n$ repetitions of
ones and zeros, respectively. The hypothesis testing rules $\hat \theta_{H,p}$
and $\hat \theta_{H,1-p}$ are chosen as described in Section~\ref{sec:control}.

\begin{figure}[!th]
  \centering
  \includegraphics{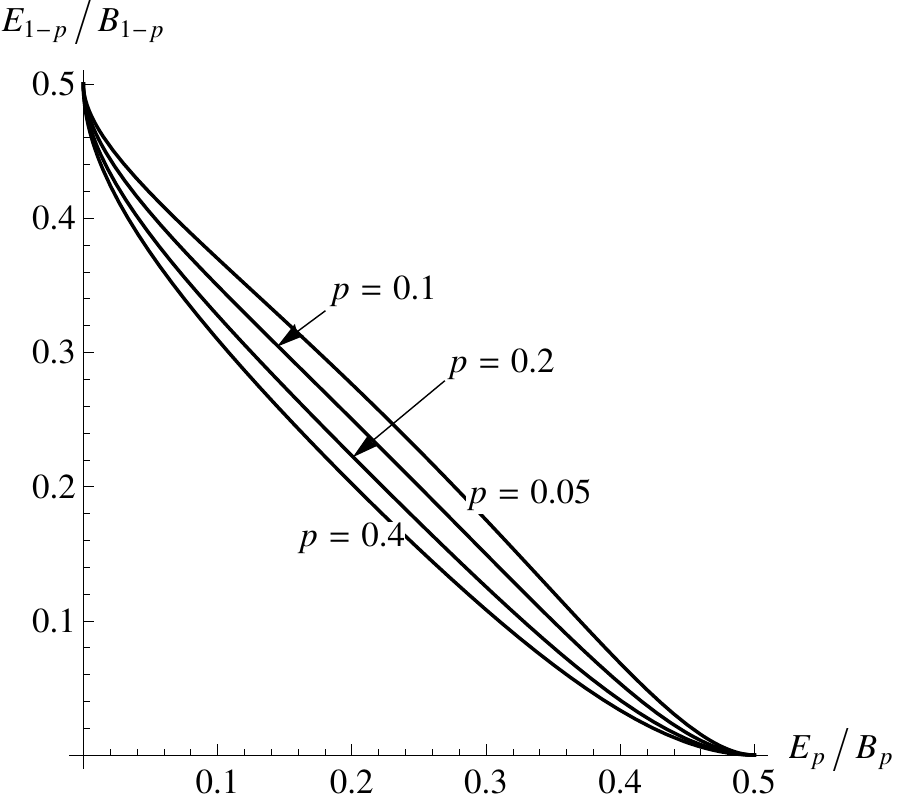}
  \caption{The scaled EER region $\Phi(p) = \{ (E_p/B_p, E_{1-p}/B_{1-p}) :
  (E_p, E_{1-p}) \in \ALPHABET E(R_p, R_{1-p}) \}$ for different instances of
  the compound channel $\ALPHABET Q_p \DEFINED \{\BSC_p, \BSC_{1-p} \}$.}
  \label{fig:EER} 
\end{figure}

Proposition~\ref{prop:exponent} implies that for any rate vector $(R_p,
R_{1-p})$ and a particular choice of the estimation threshold $q_c$,%
\footnote{The choice of $q_m$ does not affect the values of $E_p$ and $E_{1-p}$ as long as
$\PR_\ell(\hat L_m \neq \ell) \to 0$. For that, we require only that $p < q_m < 1 -p$. Choosing $q_m = 0.5$
ensures that.}
the above scheme
achieves an error exponent $(E_p, E_{1-p})$ such that
\begin{align*}
  E_p &\ge
  \frac{D(q_c \| p)   D(p \| 1-p)}{D(q_c\|p) + D(p \| 1-p)} (1-\gamma_p),\\
  E_{1-p} &\ge
  \frac{D(q_c \| 1-p) D(p \| 1-p)}{D(q_c\|1-p) + D(p \| 1-p)} (1-\gamma_{1-p})
\end{align*}
where $\gamma_p = R_p/C_p$ and $\gamma_{1-p} = R_{1-p}/C_{1-p}$. 

There are no known upper bounds on the EER. Hence, we compare with the trivial
upper bound of the Burnashev exponent of $\BSC_p$ and $\BSC_{1-p}$. 
\begin{align*}
  \frac {E_p}{B_p} &\ge 
  \frac{D(q_c \| p)}{D(q_c\|p) + D(p \| 1-p)} ,\\
  \frac {E_{1-p}}{B_{1-p}} &\ge 
  \frac{D(q_c \| 1-p)}{D(q_c\|1-p) + D(p \| 1-p)}.
\end{align*}
Let 
\begin{equation*} \label{eq:ratio}
 \varphi(p,q_c) \DEFINED \left( \frac {E_p}{B_p}, \frac{E_{1-p}}{B_{1-p}} \right) \ge
  \left( \frac{D(q_c \|p)} {D(q_c\|p) + D(p\|1-p)}, 
         \frac{D(q_c \|1-p)} {D(q_c\|1-p) + D(p\|1-p)} \right)
\end{equation*}
and
\begin{align*}
  \Phi(p) &\DEFINED \left\{ \left( \frac{E_p}{B_p}, \frac{E_{1-p}}{B_{1-p}} \right) :
  (E_p, E_{1-p}) \in \ALPHABET E(R_p, R_{1-p}) \right\} \\
  &=\{ \varphi(p,q_c) : p < q_c < p \}
\end{align*}
For the scheme proposed in Section~\ref{sec:scheme}, $\Phi(p)$ does not depend
on the transmission rate $(R_p, R_{1-p})$. We plot $\Phi(p)$ for different
values of $p$ in Figure~\ref{fig:EER}. 

\section {Conclusion} \label{sec:conclusion}

In the presence of feedback, not knowing the exact channel transition matrix does
not result in a loss in capacity. As a result, we can provide an
optimistic rate guarantee: any rate less than the capacity of the realized
channel is opportunistically achievable, even though we do not know the realized channel before the start of
communication. This is in contrast to the pessimistic rate guarantees in
compound channel without feedback. More importantly, any rate vector in the
optimistic capacity region can be achieved using a simple, training-based coding
scheme. The error exponent of this scheme has a negative slope at all rates in
the capacity region, even at rates near the boundary of the capacity region. 

Our proposed proposed training based
scheme is conceptually similar to Yamamoto-Itoh's scheme. It operates in
multiple epochs; each epoch is divided into a message mode and a control mode. A
training sequence is transmitted at the beginning of each mode, and the
corresponding channel estimate determines the operation during the remainder of
the mode.

It may appear that the proposed scheme can be simplified by combining the
training phases in each epoch, \emph{i.e.}, have a training phase followed by
message and control modes. However, as argued by Tchamkerten and
Telatar in~\cite{TchamkertenTelatar:2006}, such a simplification will lead to error
exponents that have zero-slope near capacity. Our results do not contradict the
results of~\cite{TchamkertenTelatar:2006} because we allow for more
sophisticated training. Re-training in the control mode ensures
that the error events $\{\hat W_m(k) \neq W_m(k) \}$ and $\{\hat W_c(k) \neq
W_c(k) \}$ are independent, which, in turn, is essential to obtain an error
exponent of the form $B_\ell (1 - \gamma_\ell)$. 

One possible way to make the scheme more efficient is to accumulate the
training sequences for each phase, \emph{i.e.}, the channel estimation for the
message mode and the control mode is based on all past training
sequences for that mode. Such an accumulation will improve the finite length
performance of the scheme, but does not affect the asymptotic performance
because, in the limit, the communication lasts for only one epoch with high
probability. 


Another possibility to improve the performance of the coding scheme is to use a
universal coding scheme for the control mode rather than a training based
scheme. This motivates the study of the following communication problem.

\begin{problem} \label{prob:open-prob}
  Consider the communication of a binary valued message over a compound channel
  with feedback. Let $\ALPHABET Q = \SEQ Q$ denote the compound
  channel, $W \in \{\theta_0, \theta_1\}$ denote the message, $X_t$ and $Y_t$
  denote the channel inputs and output at time $t$, and $\hat W$ denotes the
  decoded message. Consider a variable length coding scheme $(\SET c, \SET g,
  \tau)$, where $c_t$ is the encoding function at time $t$, $g_t$ is the
  decoding function at time $t$, and $\tau$ is a $Y^t$-measurable stopping time.
  The decoded message is
  \[ \hat W = g_\tau(Y_1, \dots, Y_\tau). \]

  Let $a_{\ell n}$ and $b_{\ell n}$ denote the exponent of the two types of
  errors, \emph{i.e.},
  \begin{align}
    a_{\ell n}(\SET c, \SET g, \tau) &= 
    \frac{- \log \PR_\ell(\hat W = \theta_1 \mid W = \theta_0) }
    { \EXP_\ell[\tau \mid W = \theta_0] },
    \quad \ell = 1,\dots, L, \\ 
    b_{\ell n}(\SET c, \SET g, \tau) &= 
    \frac{- \log \PR_\ell(\hat W = \theta_0 \mid W = \theta_1) }
    { \EXP_\ell[\tau \mid W = \theta_1] },
    \quad \ell = 1,\dots, L. 
  \end{align}
  where $\PR_\ell$ is the induced probability measure when the true channel
  equals $Q_\ell$. 

  For a sequence $S = \{ \SET c^{(n)}, \SET g^{(n)}, \tau^{(n)} \}_{n=1}^\infty$
  of coding schemes such that
  \[ \lim_{n \to \infty} \EXP_\ell[\tau^{(n)} \mid W = \theta_i] = \infty, \quad
  i = 0,1, \quad \ell = 1,\dots,L.
  \]
  define the type-I and type-II error exponents of $S$ as
  \begin{align*}
    a_{\ell*} &= \lim_{n \to \infty} a_{\ell n}(\SET c^{(n)}, \SET g^{(n)},
    \tau^{(n)}), \\
    b_{\ell*} &= \lim_{n \to \infty} b_{\ell n}(\SET c^{(n)}, \SET g^{(n)},
    \tau^{(n)}).
  \end{align*}
  Furthermore, define 
  \[ b^*_\ell = \max_{S : a_{\ell*}(S) = 0} b_{\ell*}(S). \]

  What is the best type-II exponent $\VEC {b^*}$?
\end{problem}
Tchamkerten and Telatar~\cite{TchamkertenTelatar:2005} studied a similar
problem and identified necessary and sufficient conditions under which
\[ b^*_\ell = B_{Q_\ell}, \quad \ell = 1, \dots, L. \]
We are not aware of the solution to the above problem when the conditions
of~\cite{TchamkertenTelatar:2005} are not satisfied. 

Given any sequence $S$ of coding schemes for Problem~\ref{prob:open-prob}, we
can replace the control mode (phases three and four) of the proposed coding
scheme by $S$ and achieve an error exponent of
\[  (b_{1*}(S)(1-\gamma_1), \dots, b_{L*}(S)(1-\gamma_L)). \]
If $S$ is optimal, the error exponent is
\begin{equation}\label{eq:pareto}
  (b^*_{1}(S)(1-\gamma_1), \dots, b^*_{L}(S)(1-\gamma_L)). 
\end{equation}
\emph{We conjecture that no coding scheme can achieve a better error exponent,
\emph{i.e.}, \eqref{eq:pareto} is the Pareto frontier of the EER.}

When the conditions of~\cite{TchamkertenTelatar:2005} are
satisfied, we can replace the control mode by the variable length coding scheme
proposed in~\cite{TchamkertenTelatar:2005}, and thereby recover the result
of~\cite{TchamkertenTelatar:2006a}. In fact, in that case, our modified scheme
is exactly the same as the variation proposed
in~\cite[Section~IV-B]{TchamkertenTelatar:2006a}. When the conditions
of~\cite{TchamkertenTelatar:2005} are not satisfied, the scheme proposed in this
paper provide an inner bound on the error exponent region. To find the best error
exponents, we need to solve Problem~1. 

In this paper, we presented an inner bound on the EER when the compound channel
is defined over a finite family. Generalization of the coding scheme to compound
channels defined over continuous families is an important and interesting
future direction. We believe that solving Problem~\ref{prob:open-prob} is a
critical step in that direction. 

\section*{Acknowledgment}

The authors are grateful to A.\ Tchamkerten for helpful feedback and to the
anonymous reviewers whose suggestions helped to improve the presentation of the
paper.

\bibliographystyle{IEEEtran}
\bibliography{IEEEabrv,../../collection}

\begin{thebibliography}{10}
\providecommand{\url}[1]{#1}
\csname url@samestyle\endcsname
\providecommand{\newblock}{\relax}
\providecommand{\bibinfo}[2]{#2}
\providecommand{\BIBentrySTDinterwordspacing}{\spaceskip=0pt\relax}
\providecommand{\BIBentryALTinterwordstretchfactor}{4}
\providecommand{\BIBentryALTinterwordspacing}{\spaceskip=\fontdimen2\font plus
\BIBentryALTinterwordstretchfactor\fontdimen3\font minus
  \fontdimen4\font\relax}
\providecommand{\BIBforeignlanguage}[2]{{%
\expandafter\ifx\csname l@#1\endcsname\relax
\typeout{** WARNING: IEEEtran.bst: No hyphenation pattern has been}%
\typeout{** loaded for the language `#1'. Using the pattern for}%
\typeout{** the default language instead.}%
\else
\language=\csname l@#1\endcsname
\fi
#2}}
\providecommand{\BIBdecl}{\relax}
\BIBdecl

\bibitem{Wolfowitz:1959}
J.~Wolfowitz, ``Simultaneous channels,'' \emph{Archive for Rational Mechanics
  and Analysis}, pp. 371--386, Nov. 1959.

\bibitem{BlackwellBreimanThomasian:1959}
D.~Blackwell, L.~Breiman, and A.~J. Thomasian, ``The capacity of a class of
  channels,'' \emph{The Annals of Mathematical Statistics}, vol.~30, no.~4, pp.
  1229--1241, Dec. 1959.

\bibitem{Wolfowitz:1964}
J.~Wolfowitz, \emph{Coding Theorems of Information Theory}.\hskip 1em plus
  0.5em minus 0.4em\relax Springer Verlag, 1964.

\bibitem{ShraderPermuter:2009}
B.~Shrader and H.~Permuter, ``Feedback capacity of the compound channel,''
  \emph{{IEEE} Trans. Inf. Theory}, vol.~55, pp. 3629--3644, Aug. 2009.

\bibitem{LapidothNarayan:1998}
A.~Lapidoth and P.~Narayan, ``Reliable communication under channel
  uncertainty,'' \emph{{IEEE} Trans. Inf. Theory}, vol.~44, no.~6, pp.
  2148--2177, Oct. 1998.

\bibitem{Burnashev:1976}
M.~V. Burnashev, ``Data transmission over a discrete channel with feedback.
  {R}andom transmission time,'' \emph{Problemy peredachi informat︠s︡ii},
  vol.~12, no.~4, pp. 10--30, 1976.

\bibitem{TchamkertenTelatar:2006a}
A.~Tchamkerten and I.~E. Telatar, ``Variable length coding over an unknown
  channel,'' \emph{{IEEE} Trans. Inf. Theory}, vol.~52, no.~5, pp. 2126--2145,
  May 2006.

\bibitem{WengPradhanAnastasopoulos:2008}
L.~Weng, S.~S. Pradhan, and A.~Anastasopoulos, ``Error exponent regions for
  {G}aussian broadcast and multiple-access channels,'' \emph{{IEEE} Trans. Inf.
  Theory}, vol.~54, no.~7, pp. 2919--2942, Jul. 2008.

\bibitem{ByersLubyMitzenmacherRege:1998}
J.~W. Byers, M.~Luby, M.~Mitzenmacher, and A.~Rege, ``A digital fountain
  approach to reliable distribution of bulk data,'' \emph{SIGCOMM Comput.
  Commun. Rev.}, vol.~28, no.~4, pp. 56--67, 1998.

\bibitem{Luby:2002}
M.~Luby, ``{LT} codes,'' in \emph{Proceedings of the IEEE Symposium on
  Foundations of Computer Science}, 2002, pp. 271--280.

\bibitem{Shokrollahi:2006}
A.~Shokrollahi, ``Raptor codes,'' \emph{{IEEE} Trans. Inf. Theory}, vol.~6,
  no.~52, pp. 2551--2567, Jun. 2006.

\bibitem{YamamotoItoh:1979}
H.~Yamamoto and K.~Itoh, ``Asymptotic performance of a modified
  {S}chalkwijk-{B}arron scheme for channels with noiseless feedback,''
  \emph{{IEEE} Trans. Inf. Theory}, vol.~25, no.~6, pp. 729--733, Nov. 1979.

\bibitem{Tuncel:2005}
E.~Tuncel, ``On error exponents in hypothesis testing,'' \emph{{IEEE} Trans.
  Inf. Theory}, vol.~51, no.~8, pp. 2945--2950, Aug. 2005.

\bibitem{Blahut:1974}
R.~E. Blahut, ``Hypothesis testing and information theory,'' \emph{{IEEE}
  Trans. Inf. Theory}, no.~4, Jul. 1974.

\bibitem{Hoeffding:1963}
W.~Hoeffding, ``Probability inequalities for sums of bounded random
  variables,'' \emph{Journal of the American Statistical Association}, vol.~58,
  no. 301, pp. 13--30, Mar. 1963.

\bibitem{TchamkertenTelatar:2006}
A.~Tchamkerten and I.~E. Telatar, ``On the use of training sequences for
  channel estimation,'' \emph{{IEEE} Trans. Inf. Theory}, vol.~52, no.~3, pp.
  1171--1176, Mar. 2006.

\bibitem{TchamkertenTelatar:2005}
------, ``On the universality of {B}urnashev's exponent,'' \emph{{IEEE} Trans.
  Inf. Theory}, vol.~51, no.~8, pp. 2940--2944, Aug. 2005.

\end{thebibliography}

\end{document}